\documentclass[10pt]{article}

\usepackage{graphicx}

\begin{document}

\title{Some anisotropic universes in the presence of imperfect fluid coupling with spatial curvature}
\date{}
\author{\"{O}zg\"{u}r Akarsu\footnote{Department of Physics, Ko\c{c} University, 34450 Sar{\i}yer, \.{I}stanbul, Turkey. \textbf{E-Mail:} oakarsu@ku.edu.tr} \and Can Battal K{\i}l{\i}n\c{c}\footnote{Department of Astronomy and Space Sciences, Ege University, 35100 Bornova, {\.I}zmir, Turkey. \textbf{E-Mail:} can.kilinc@ege.edu.tr}}
\maketitle
\begin{center}
\vskip-1cm
\textit{}
\end{center}

\begin{abstract}
We consider Bianchi VI spacetime, which also can be reduced to Bianchi types $\rm{VI_{0}}$-V-III-I. We initially consider the most general form of the energy-momentum tensor which yields anisotropic stress and heat flow. We then derive an energy-momentum tensor that couples with the spatial curvature in a way so as to cancel out the terms that arise due to the spatial curvature in the evolution equations of the Einstein field equations. We obtain exact solutions for the universes indefinetly expanding with constant mean deceleration parameter. The solutions are beriefly discussed for each Bianchi type. The dynamics of the models and fluid are examined briefly, and the models that can approach to isotropy are determined. We conclude that even if the observed universe is almost isotropic, this does not necessarily imply the isotropy of the fluid (e.g., dark energy) affecting the evolution of the universe within the context of general relativity.

\begin{flushleft}
\end{flushleft}

\end{abstract}
\section{Introduction}
\label{intro}

The Bianchi universes form almost complete class of spatially homogenous but not necessarily isotropic relativistic cosmological models. They provide generalizations of the standard Friedmann-Lema\^{i}tre (FL) models, which are based on the spatially homogeneous and isotropic Robertson-Walker (RW) metrics \cite{Ellis06,Ellis08}. Such models are of interest in cosmology in favor of constructing more realistic models than the FLRW models with maximally symmetric spatial geometry. Namely, although the observed universe seems to be almost isotropic on large scales, the early and/or very late universe could be anisotropic \cite{Ellis06}. Additionally, the interest in such models was promoted in recent years due to the debate that going on the analysis and the interpretation of the Wilkinson Microwave Anisotropy Probe (WMAP) \cite{Komatsu10,Komatsu09,Spergel07,Peiris03} data, whether they need a Bianchi type morphology to be explained successfully \cite{Bennett03,Oliveira,Schwarz04,Cruz,Hoftuft09,BennettHill}.

While the spatial isotropy of RW metrics exclude the anisotropic stress and/or energy flux, once the metric is generalized to Bianchi types, perfect fluid representantion of the energy-momentum tensor (EMT) can also be generalized in accordance with the considered Bianchi type; e.g., the EMT can yield off-diagonal terms and anisotropic stress. However, the investigation of the anisotropic stresses has been widely neglected until recent years \cite{Koivisto06}. That might be because of the expectation that a dominance of an anisotropic stress would give rise to an anisotropic expansion whereas the observed universe appears to expand isotropically. However, the anisotropic stresses have been considered by some authors. For instance, Barrow \cite{Barrow97} analyzed the cosmological evolution of matter sources that possess anisotropic pressures (by assuming the anisotropy is very small) such as electric and magnetic fields, collisionless relativistic particles, spatial curvature anisotropies, etc. Besides these known sources, since the release of WMAP data, the possible inflaton field and dark energy candidates that yield anisotropic pressure have been gaining increasing interest in favor of resolving the likely anomalies in WMAP data. Several cosmological models which introduce a source that yield anisotropic stress to break the spatial isotropy, either during inflation or in the late time acceleration of the universe, have been proposed, e.g., by the anisotropic inflaton field (say, vectorial fields) \cite{Golovnev,Koivisto08a}, by the anisotropic cosmological constant or dark energy \cite{Koivisto08a,Rodrigues,Koivisto08b,Koivisto08c,AkarsuLRS,AkarsuSitter,AkarsuBIII,SharifZubair10a,SharifZubair10b,Yadav10a}. However, even if it turns out that these anomalies are statisticaly insignificant and standard $\Lambda$ cold dark matter model fits the WMAP data very well, we still can not rule out such energy sources. The reason being that anisotropic stresses do not necessarily promote anisotropy in the expansion, and there can even exist such anisotropic stresses that can cause universe to isotropize as Bianchi type I (or type V) universes in the presence of isotropic fluid (\cite{AkarsuBIII}).

The spatial curvature is referred as a possible source of anisotropic stresses in general relativistic cosmologies by Barrow \cite{Barrow97,Barrow95}. According to this idea, the three-curvature terms in the Einstein tensor can be moved to the other side of the Einstein equations and reinterpreted as an additional "effective" EMT, and then the "effective" EMT can behave as an anisotropic source. Thus, any Bianchi type, with anisotropic spatial curvature, universe containing an isotropic fluid, can behave like a Bianchi type I universe containing that fluid plus an additional anisotropic fluid (which obtained by the method mentioned above). On the other hand, one may also set an interesting class of solutions by following a similar idea. One may assume an EMT which compensates the three-curvature terms that moved to the other side of the Einstein equations and the total effective EMT becomes isotropic whether the metric possesses anisotropic spatial curvature or not. Hence, in that case, any Bianchi type, with anisotropic spatial curvature, universe containing this EMT, can behave like a Bianchi type I universe containing an isotropic fluid. Accordingly, one may assume a fluid that couples with the spatial curvature of the universe in a way so as to vanish the terms arise due to the deviation from spatially flat geometry in the evolution equations. Once that is assumed any anisotropy in the spatial curvature will give rise to an anisotropic stress in the fluid. There are no a priori nor observational reasons to exclude such a fluid from being candidate for the inflaton field and/or dark energy, whose natures are not clear yet. Conversely, such an approach can provide interesting generalizations of the known candidates for inflaton field and/or dark energy.

This study based on the above idea  is carried out in the framework of Bianchi type VI spacetime, which can be reduced to the Bianchi types $\rm{VI_{0}}$, III with anisotropic spatial curvature and types V, I with isotropic spatial curvature. Compared the Bianchi types I \cite{SharifZubair10a,KumarSingh} and V \cite{SinghT}, Bianchi types III \cite{Yadav10a,Behera10,Tiwari09,SinghJP07} and $\rm{VI_{0}}$ \cite{SharifZubair10b,Roy85,Ram88,Coley90,Bali,Tripathy} spacetimes have been studied rarely in literature of general relativistic cosmology. The basic reason being that these spacetimes do not cover RW spacetimes. However, they can still approach arbitrarily close to RW spacetimes. Thus, these spacetimes can still be studied for constructing realistic cosmological models.

We, initially, do not constraint the EMT and give full freedom to its components conveniently with the considered metric; hence, it yields anisotropic pressure and heat flow. We do not ignore the heat flow basically in favor of not compromising from generality. Heat flow can be important in the early stages of the universe and has been studied by several authors, e.g., \cite{Nayak89,Roy96,SinghZeRam,Singh09a,Singh09b}. Beside that, the heat flow can give rise to a temperature gradient along a certain axis of the universe. Hence, it would also be necessary to examine whether it tends to zero in the late universe in our models, otherwise the universe would not appear isotropic.

We obtain exact solutions for the universes that exhibit volumetric expansion with constant mean deceleration parameter. Such solutions are not inconsistent with observations, can also be appraised approximately valid for slowly time varying deceleration parameter and have widely been considered (e.g., \cite{SharifZubair10b,Yadav10a,KumarSingh,Singh07,Singh08,AdhavBansod,YadavRahaman,PradhanJotania}). Such cosmological models have also been studied in the presence of various anisotropic fluids by several authors recently.  Sharif and Zubair investigated Bianchi $\textnormal{VI}_{0}$ models in the presence of cosmological constant, anisotropic dark energy and electromagnetic field \cite{SharifZubair10b}. Yadav and Yadav investigated a Bianchi III model in the presence of anisotropic dark energy by assuming the expansion scalar is proportional to shear scalar \cite{Yadav10a}. Akarsu and Kilinc \cite{AkarsuLRS} investigated LRS Bianchi I models in the presence of perfect fluid and minimally interacting anisotropic dark energy. The same authors \cite{AkarsuBIII} presented Bianchi III models, which can isotropize as the Bianchi type I models in the presence of isotropic fluid, in the presence of single anisotropic fluid. They \cite{AkarsuSitter} also discussed the de Sitter volumetric expansion in the presence of anisotropic fluid in detail within the framework of Bianchi I spacetime and presented two models in the presence of an anisotropic fluid that obtained by minimally altering the conventional vacuum energy.

This paper is constructed as follows: In Sec. 2, the metric and some basic kinematical equations are introduced. In Sec. 3, we discuss some basic dynamics of the models in the context of general relativity and derive an EMT that couples with the spatial curvature in a way so as to cancel out the terms that arise due to the spatial curvature in the evolution equations. In Sec. 4, we obtain exact solutions for the universes that exhibit volumetric expansion with constant mean deceleration parameter. We briefly discuss the kinematics and dynamics of the models, and determine the models that can approach to isotropy.

\section{The metric and some basic kinematical equations}
\label{sec:1}
For convenience, we use the natural units, i.e., $8\pi G=1$ and $c=1$. The Ricci tensor is defined as the contraction of the Riemann tensor as follows: $R_{\mu\nu}={R^{\lambda}}_{\mu\nu\lambda}$, where $\mu$, $\nu$ and $\lambda$ run from 0 to 3. Similarly, for the spatial section of the metric, the three-Ricci tensor is defined as follows: $^3R_{ij}={^3R^{k}}_{ijk}$, where $i$, $j$ and $k$ run over spatial components from 1 to 3.

We consider the spatially homogenous but anisotropic spacetime described by Bianchi type VI metric in the form
\begin{equation}
ds^{2}=-dt^{2}+A(t)^{2}e^{-2mz}dx^{2}+B(t)^{2}e^{2nz}dy^{2}+C(t)^{2}dz^{2},
\end{equation}
where $m$ and $n$ are real constants, and $A$, $B$ and $C$ are functions of the cosmic time $t$ only. It can be observed that the metric (1) can be reduced to Bianchi type $\rm{VI_{0}}$ (B$\rm{VI_{0}}$) when $n=m\neq 0$, Bianchi type III (BIII) when $n=0\neq m$, Bianchi type V (BV) when $n=-m\neq 0$ and Bianchi type I (BI) when $n=m=0$. While BI covers the spatially flat RW spacetime, BV covers the spatially open RW spacetime. On the other hand, the underlying Lie algebra of the isometry group of B$\rm{VI_{0}}$ and BIII are completely different from those of BV and BI, and they do not cover RW spacetimes. However, they can evolve to a phase where they are observationally indistinguishable from RW spacetimes.

Non-zero components of the three-Ricci tensors ($^{3}R^{i}_{j}$) and the three-Ricci scalar ($^{3}R={^{3}R^{i}_{i}}$), which shall be used in the derivation of the energy-mometum tensor that would be considered in the solutions, of the metric (1) are as follows:
\begin{eqnarray}
{^{3}R^{1}_{1}}=\frac{m^2-mn}{C^{2}},\quad
{^{3}R^{2}_{2}}=\frac{n^2-mn}{C^{2}}\quad\textnormal{and}\quad
{^{3}R^{3}_{3}}=\frac{m^2+n^2}{C^{2}}
\end{eqnarray}
and
\begin{equation}
^{3}R=\frac{2m^{2}-2mn+2n^{2}}{C^{2}}.
\end{equation}
It can be observed that the three-Ricci tensors are not all identical unless $n=-m$. Strictly speaking, the metric has an anisotropic spatial curvature when $n\neq -m$ (BIII and B$\rm{VI_{0}}$ cases), but has an isotropic spatial curvature when $n=-m$ (BI and BV cases). The spatial curvature parameters, $m$ and $n$, shall appear as the constants of the motion in the evolution equations (see Eqs. (14)-(16)). Accordingly, the flatness of the space cannot change as the universe expands independent of whether the universe undergoes an accelerated expansion period or not. However, an accelerated period of expansion (strictly speaking, in the $z$ axis, i.e., $C$, for the metric we consider) can drive the three-Ricci tensors and three-Ricci scalar indistinguishably close to zero. In other words, the universe can evolve to a phase where it is observationally indistinguishable from spatially flat geometry.

The directional Hubble parameters, which express the expansion rates of the universe in the directions of $x$, $y$ and $z$ respectively, can be defined as follows:
\begin{equation}
H_{1}=\frac{\dot{A}}{A}\textnormal{,}\qquad H_{2}=\frac{\dot{B}}{B}\qquad \textnormal{and}\qquad H_{3}=\frac{\dot{C}}{C},
\end{equation}
where an overdot denotes $d/dt$. The mean Hubble parameter, which expresses the volumetric expansion rate of the universe, can also be given as 
\begin{equation}
H=\frac{1}{3}\frac{\dot{V}}{V}=\frac{1}{3}\sum_{i=1}^{3}H_{i},
\end{equation}
where $V=ABC$ is the volume scale factor of the universe.

Two other kinematical quantities of observational interest in cosmology are the anisotropy parameter of the expansion $\Delta$ and the dimensionless mean deceleration parameter $q$, which are defined as
\begin{equation}
\Delta=\frac{2}{3}\frac{\sigma^{2}}{H^{2}}=\frac{1}{3}\sum_{i=1}^{3}\left(\frac{H_{i}-H}{H}\right)^{2},
\end{equation}
\begin{equation}
q=\frac{d}{dt}\left(\frac{1}{H}\right)-1,
\end{equation}
where ${\sigma}^{2}=\frac{1}{2}\sigma_{ij}\sigma^{ij}$ ($\sigma_{ij}$ is the shear tensor) is the shear scalar. $\Delta$ is the measure of the deviation from isotropic expansion and the universe expands isotropically if $\Delta=0$. The universe exhibits accelerating volumetric expansion if $-1\leq q<0$, decelerating volumetric expansion if $q>0$, and exhibits constant-rate volumetric expansion if $q=0$.

\section{Some basic dynamics and derivation of the energy-momentum tensor}
\label{sec:2}
We assume that the fluid four-velocity is comoving; thus $u^{\mu}=\delta_{0}^{\mu}$ with $u_{\mu}u^{\mu}=-1$, where $\delta_{\nu}^{\mu}$ is the unit four-tensor. The EMT of the fluid within the framework of the metric (1) can most generally be written in the following form;
\begin{equation}
{T}_{\nu}^{\mu} = \left( \begin{array}{cccc}
-\rho & 0 & 0 & h_{3} \\
0 & p_{1} & 0 & 0 \\
0 & 0 & p_{2} & 0 \\
-\frac{h_{3}}{C^2} & 0 & 0 & p_{3} \end{array} \right),
\end{equation}
where $\rho$ is the energy density of the fluid; $p_{1}$, $p_{2}$ and $p_{3}$ are the directional pressures on the $x$, $y$ and $z$ axes respectively, and $h_{3}$ is the energy flux density along the $z$ axis. Because the fluid is comoving, energy flux due to the bulk motion of the fluid is not possible. However, heat flow can contribute to the energy flux, and then, $h_{3}$ may be interpreted as the heat flow along the $z$ axis, see \cite{Landau,McGlinn}.

Because the fluid is in the rest frame, anisotropy in its pressure can arise due to the anisotropy in its equation of state (EoS) parameter, since the energy density is a scalar quantity. Accordingly, one may assume that the directional pressures of the anisotropic fluid are each proportional to its energy density, i.e.,
\begin{equation}
(p_{1},p_{2},p_{3})=(w_{1}\rho, w_{2}\rho, w_{3}\rho),
\end{equation}
where $w_{1}$, $w_{2}$ and $w_{3}$ are the directional EoS parameters on the $x$, $y$ and $z$ axes respectively and they can be functions of the cosmic time $t$. However, for brevity, we handle with pressures rather than directional EoS parameters in the stream of solutions, though we shall deal with them while discussing the physical behaviors of the models.

The EMT given in (8) may be parametrized as follows:
\begin{equation}
{T}_{\nu}^{\mu} = \left( \begin{array}{cccc}
-\rho & 0 & 0 & h_{3} \\
0 & p+\tilde{p}_{1} & 0 & 0 \\
0 & 0 & p+\tilde{p}_{2} & 0 \\
-\frac{h_{3}}{C^2} & 0 & 0 & p+\tilde{p}_{3} \end{array} \right),
\end{equation}
where $p$ is the deviation-free part of the pressure (its meaning shall be specified below), and $\tilde{p}_{1}$, $\tilde{p}_{2}$ and $\tilde{p}_{3}$ are the deviations from $p$ on the $x$, $y$ and $z$ axes respectively, i.e.,
\begin{equation}
\tilde{p}_{1}=p_{1}-p,\qquad \tilde{p}_{2}=p_{2}-p\qquad \textnormal{and} \qquad \tilde{p}_{3}=p_{3}-p.
\end{equation}

The Einstein field equations, in mixed components, can be written as
\begin{equation}
G_{\nu}^{\mu}=R_{\nu}^{\mu}-\frac{1}{2}R\delta_{\nu}^{\mu}=-T_{\nu}^{\mu},
\end{equation}
where $G_{\nu}^{\mu}$ is the Einstein tensor and $R=R^{\mu}_{\mu}$ is the Ricci scalar.

In a comoving coordinate system, Einstein field equations (12) within the framework of the metric (1), in case of (10), lead to the following system of equations:
\begin{equation}
\frac{\dot{A}}{A}\frac{\dot{B}}{B}+\frac{\dot{A}}{A}\frac{\dot{C}}{C}+\frac{\dot{B}}{B}\frac{\dot{C}}{C}-\frac{m^2-mn+n^2}{C^{2}}=\rho,
\end{equation}
\begin{equation}
\frac{\ddot{B}}{B}+\frac{\ddot{C}}{C}+\frac{\dot{B}}{B}\frac{\dot{C}}{C}-\frac{n^2}{C^{2}}=-(p+\tilde{p}_{1}),
\end{equation}
\begin{equation}
\frac{\ddot{A}}{A}+\frac{\ddot{C}}{C}+\frac{\dot{A}}{A}\frac{\dot{C}}{C}-\frac{m^2}{C^{2}}=-(p+\tilde{p}_{2}),
\end{equation}
\begin{equation}
\frac{\ddot{A}}{A}+\frac{\ddot{B}}{B}+\frac{\dot{A}}{A}\frac{\dot{B}}{B}+\frac{mn}{C^{2}}=-(p+\tilde{p}_{3}),
\end{equation}
\begin{equation}
m\left(\frac{\dot{C}}{C}-\frac{\dot{A}}{A}\right)+n\left(\frac{\dot{B}}{B}-\frac{\dot{C}}{C}\right)=h_{3}.
\end{equation}

The conservation of the EMT tensor given in (10), ${T^{\mu\nu}}_{;\nu}=0$, leads to the following two equations;
\begin{eqnarray}
\nonumber
\dot{\rho}+3(\rho +p)H+\tilde{p}_{1}H_{1}+\tilde{p}_{2}H_{2}+\tilde{p}_{3}H_{3}=\frac{m-n}{C^{2}}h_{3},\\
\dot{h}+3hH-m(\tilde{p}_{3}-\tilde{p}_{1})+n(\tilde{p}_{3}-\tilde{p}_{2})=0.
\end{eqnarray}
In fact, from twice contracted Bianchi identity we have ${G^{\mu\nu}}_{;\nu}=0$ for the Einstein tensor and thus considering (12) one may observe that Bianchi identity already assures the conservation law for the EMT given in (10) in general relativity. One may check that the solutions presented in the following sections satisfy both of these equations.

The energy density of the fluid can be written in terms of the geometrical parameters, by using the constraint equation (13) and the definitions of $\Delta$ and $H$, as follows:
\begin{equation}
\rho=3H^{2} \left(1-\frac{\Delta}{2}\right)-\frac{m^{2}-mn+n^{2}}{C^{2}}.
\end{equation}

The sum of the directional pressures can also be written in terms of the geometrical parameters, by using (13)-(16), (19) and the definitions of $\Delta$, $H$ and $q$, as follows:
\begin{equation}
3p+\tilde{p}_{1}+\tilde{p}_{2}+\tilde{p}_{3}=3H^{2}\left(-1+2q-\frac{3}{2}\Delta\right)+\frac{m^{2}-mn+n^{2}}{C^{2}}.
\end{equation}

The behavior of $\Delta$, which appears in both (19) and (20), is crucial for deciding whether the models approach isotropy or not. According to the conditions given by Collins and Hawking \cite{CollinsHawking}, the model approaches isotropy if as $t\rightarrow \infty$
\begin{eqnarray}
\nonumber
\textnormal{i)}\;V\rightarrow\infty,\quad
\textnormal{ii)}\;\Delta\rightarrow 0
\quad\textnormal{and}\quad\
\textnormal{iii)}\;\rho>0\quad\textnormal{and}\quad\frac{h_{3}}{\rho}\rightarrow 0.
\end{eqnarray}
The first two of these conditions are geometrical constrains and the spatial section of the metric approaches isotropy if both of them are satisfied. Considering the physical ingredient of the models, we shall say that the models approach isotropy when the last condition (iii) is satisfied as well as (i) and (ii). The condition $\frac{h_{3}}{\rho}\rightarrow 0$ is taken because otherwise there would be a temperature gradient along the $z$ axis and the universe would not appear isotropic.

To express $\Delta$ explicitly we first obtain the differences between the directional Hubble parameters by using the evolution equations (14)-(16);
\begin{eqnarray}
H_{b}-H_{a}=\frac{1}{V}\left[\lambda_{b+a}+\int\left(\tilde{p}_{b}-\tilde{p}_{a}+{^{3}R^{b}_{b}}-{^{3}R^{a}_{a}}\right)Vdt\right]\;\textnormal{and} \; a>b,
\end{eqnarray}
where $a$ and $b$ run over spatial components and summation convention is not applied for $a$ and $b$ indices, and $\lambda_{3}$, $\lambda_{5}$ and $\lambda_{4}=\lambda_{3}+\lambda_{5}$ are real constants of integration. Now using this equation together with $H$ and the three-Ricci tensors (2), in the definition of $\Delta$ we obtain the anisotropy parameter of the expansion explicitly as follows:
\begin{eqnarray}
\Delta=\frac{1}{9}\frac{1}{H^{2}V^{2}}\sum_{a,b=1}^{3}\left[\lambda_{a+b}+\int\left(\tilde{p}_{b}-\tilde{p}_{a}+{^{3}R^{b}_{b}}-{^{3}R^{a}_{a}}\right)Vdt\right]^{2}\;\textnormal{and} \; a>b.
\end{eqnarray}
One can observe that the integral term in (22), contains both the differences of the deviations from $p$ (i.e, $\tilde{p}_{b}-\tilde{p}_{a}$) and the differences of the three-Ricci tensors (i.e., ${^{3}R^{b}_{b}}-{^{3}R^{a}_{a}}$). Thus, trivially, in the presence of an isotropic fluid ($\tilde{p}_{1}=\tilde{p}_{2}=\tilde{p}_{3}$) and isotropic spatial curvature (${^{3}R^{1}_{1}}={^{3}R^{2}_{2}}={^{3}R^{3}_{3}}$) the integral term vanishes and the anisotropy parameter of the expansion is reduced to the following well known simple form (e.g., see \cite{Gron}):
\begin{eqnarray}
\Delta=\frac{K^{2}}{H^{2}V^{2}},
\end{eqnarray}
where $K^{2}(=\frac{1}{9}\sum_{a,b=1}^{3}\left[\lambda_{a+b}\right]^{2}\;\textnormal{and} \; a>b)$ is constant.

However, this is not the only option that permits the reduction of the general form of the anisotropy parameter (22) to this simple form (23). There is yet another option, which was first proposed by Akarsu and Kilinc \cite{AkarsuBIII}, but without unveiling its relation with spatial curvature. One may allow coupling of the spatial curvature with the fluid, i.e., the content of the integral term may be interpreted as $\tilde{p}_{b}+{^{3}R^{b}_{b}}$ and $\tilde{p}_{a}+{^{3}R^{a}_{a}}$. In convenience with this interpretation, we may assume that the deviations from $p$ (i.e., $\tilde{p}_{1}$, $\tilde{p}_{2}$ and $\tilde{p}_{3}$) originate from the spatial curvature in a way so as to cancel out the terms arise due to the spatial curvature in the evolution equations (14)-(16), i.e.,
\begin{eqnarray}
\tilde{p}_{1}+\:{^{3}R^{1}_{1}}=\tilde{p}_{2}+\:{^{3}R^{2}_{2}}=\tilde{p}_{3}+\:{^{3}R^{3}_{3}}=\frac{1}{2}\:{^3R}
\end{eqnarray}
and so that, using the three-Ricci tensors and three-scalar, we have
\begin{eqnarray}
\tilde{p}_{1}=\frac{n^{2}}{C^{2}},\qquad \tilde{p}_{2}=\frac{m^{2}}{C^{2}},\qquad\textnormal{and}\qquad\tilde{p}_{3}=-\frac{mn}{C^{2}}.
\end{eqnarray}
Doing that, as would be seen in the following section, the terms arise due to the spatial curvature of the metric disappear in the evolution equations. Now, finally, substituting (25) in (10) we obtain an EMT in the following form:
\begin{equation}
{T}_{\nu}^{\mu} =
\left( \begin{array}{cccc}
-\rho & 0 & 0 & h_{3} \\
0 & p+\frac{n^{2}}{C^{2}} & 0 & 0 \\
0 & 0 & p+\frac{m^{2}}{C^{2}} & 0 \\
-\frac{h_{3}}{C^2} & 0 & 0 & p-\frac{mn}{C^{2}} \end{array} \right),
\end{equation}
which also satisfies equations (18) that arise from the conservation law of the EMT. One can observe that if the spatial curvature of the universe is isotropic (i.e., when $n=-m$), then the EMT is also isotropic. This is the case for BI and BV. On the other hand, if the spatial curvature of the universe is anisotropic (i.e., when $n\neq-m$), then the EMT is also anisotropic. This is the case for BIII and B$\rm{VI_{0}}$. One can also check that, in case of BIII ($n=0\neq m$) and when heat flow is null ($h_{3}=0$), the EMT is reduced to a form which is equivalent to the one given by Akarsu and Kilinc \cite{AkarsuBIII}.

Upon deriving the EMT, the meaning of the deviation-free part of the pressure ($p$) can also be specified. Substituting (25) in (20) we obtain
\begin{equation}
p=(-1+2q-\frac{3}{2}\Delta)H^{2}.
\end{equation}
Note that, all terms that appear in this equation are dependent only on the volumetric evolution of the universe, due to the fact that $\Delta$, $q$ and $H$ are parameters which are only dependable on the volume scale factor $V$ and its time derivatives. Thus, $p$ is the extracted part of the $p_{1}$, $p_{2}$ and $p_{3}$, which is independent of the type of the metric as long as the volume of the universe is independent of the the spatial curvature parameters ($m$ and $n$).

Below we give exact models in the presence of the EMT (26) that derived according to (25), for the universe that expands with constant mean deceleration parameter.

\section{The models}
In a comoving coordinate system, the Einstein field equations (12) within the framework of the metric (1) and with the EMT (26), reduce to
\begin{equation}
\frac{\dot{A}}{A}\frac{\dot{B}}{B}+\frac{\dot{A}}{A}\frac{\dot{C}}{C}+\frac{\dot{B}}{B}\frac{\dot{C}}{C}-\frac{m^2-mn+n^2}{C^{2}}=\rho,
\end{equation}
\begin{equation}
\frac{\ddot{B}}{B}+\frac{\ddot{C}}{C}+\frac{\dot{B}}{B}\frac{\dot{C}}{C}=-p,
\end{equation}
\begin{equation}
\frac{\ddot{A}}{A}+\frac{\ddot{C}}{C}+\frac{\dot{A}}{A}\frac{\dot{C}}{C}=-p,
\end{equation}
\begin{equation}
\frac{\ddot{A}}{A}+\frac{\ddot{B}}{B}+\frac{\dot{A}}{A}\frac{\dot{B}}{B}=-p,
\end{equation}
\begin{equation}
m\left(\frac{\dot{C}}{C}-\frac{\dot{A}}{A}\right)+n\left(\frac{\dot{B}}{B}-\frac{\dot{C}}{C}\right)=h_{3}.
\end{equation}
This system of the equations is not fully determined, because there are five linearly independent equations (28)-(32), but six variables ($A$, $B$, $C$, $\rho$, $p$ and $h_{3}$). Thus an extra constraint is necessary to close the system. Accordingly, we assume that the mean deceleration parameter $q$ is constant. Such an assumption is not inconsistent with observations \cite{KumarSingh,Singh07} and can also be appraised approximately valid for slowly time varying deceleration parameter \cite{Singh08}. One can exhaust all the possible constant $q$ values for an indefinitely expanding universe when considering the exponential volumetric expansion (33) ($q=-1$) and the power-law volumetric expansion (34) ($q>-1$);
\begin{equation}
V=c_{1}e^{3kt}
\end{equation}
and
\begin{equation}
V=a_{1}t^{3l},
\end{equation}
where $c_{1}$, $k$ and $a_{1}$, $l$ are positive constants. From (5) and (7), $q=-1$ for the exponential volumetric expansion and $q=l^{-1}-1$ for the power-law volumetric expansion. Thus, the models with the exponential volumetric expansion and power-law volumetric expansion for $l>1$ exhibit accelerating volumetric expansion. The model for $l=1$ exhibits constant-rate volumetric expansion, the models for $l<1$ exhibit decelerating volumetric expansion. Thus, one may, phenomenologically, consider the anisotropic fluid we derived in the context of inflaton field or dark energy in the models with exponential expansion and power-law expansion for $l>1$.

It can be observed that, for the volumetric expansion laws given in (33) and (34), $q$ and $H$ are independent of the spatial curvature parameters ($m$ and $n$), so do $\Delta$ and $p$ from (23) and (27) respectively. Hence, the kinematical properties of the models are independent of the spatial curvature parameters, i.e., $A$, $B$, $C$, $H$, $\Delta$ are independent of $m$ and $n$. On the other hand, the properties of the fluid ($\rho$, $p_{1}$, $p_{2}$, $p_{3}$, $h_{3}$) and, trivially, spatial curvature properties of the models ($R$, ${^{3}R^{\nu}_{\mu}}$) are dependent on the curvature parameters.

\subsection{The models for exponential volumetric expansion}
Using the evolution equations (29-31) for the exponential volumetric expansion (33), we obtain the scale factors as follows:
\begin{equation}
A=\frac{c_{1}}{c_{2}c_{3}}e^{kt-\frac{c_{4}+c_{5}}{3k}e^{-3kt}},
\end{equation}
\begin{equation}
B=c_{2}e^{kt+\frac{c_{5}}{3k}e^{-3kt}},
\end{equation}
\begin{equation}
C=c_{3}e^{kt+\frac{c_{4}}{3k}e^{-3kt}},
\end{equation}
where $c_{2}$ and $c_{3}$ are positive constants and $c_{4}$ and $c_{5}$ are real constants.

The mean Hubble parameter is,
\begin{equation}
H=k
\end{equation}
and the directional Hubble parameters on the $x$, $y$ and $z$ axes are, respectively,
\begin{equation}
H_{1}=k+(c_{4}+c_{5})e^{-3kt},
\end{equation}
\begin{equation}
H_{2}=k-c_{5}e^{-3kt},
\end{equation}
\begin{equation}
H_{3}=k-c_{4}e^{-3kt}.
\end{equation}

Using the directional and mean Hubble parameters in (6), the anisotropy of the expansion is obtained as follows:
\begin{equation}
\Delta=\frac{2}{3}\frac{{c_{4}}^{2}+c_{4}c_{5}+{c_{5}}^{2}}{k^{2}}e^{-6kt}.
\end{equation}

Using the scale factor on the $z$ axis (37) in (2) and (3), the three-Ricci tensors and the three-Ricci scalar are obtained as follows:
\begin{equation}
{^{3}R^{1}_{1}}=\frac{m^{2}-mn}{{c_{3}}^{2}}e^{-2kt-\frac{2}{3}\frac{c_{4}}{k}e^{-3kt}},
\end{equation}
\begin{equation}
{^{3}R^{2}_{2}}=\frac{n^{2}-mn}{{c_{3}}^{2}}e^{-2kt-\frac{2}{3}\frac{c_{4}}{k}e^{-3kt}},
\end{equation}
\begin{equation}
{^{3}R^{3}_{3}}=\frac{m^{2}+n^{2}}{{c_{3}}^{2}}e^{-2kt-\frac{2}{3}\frac{c_{4}}{k}e^{-3kt}},
\end{equation}
and
\begin{equation}
^{3}R=\frac{2m^{2}-2mn+2n^{2}}{{c_{3}}^{2}}e^{-2kt-\frac{2}{3}\frac{c_{4}}{k}e^{-3kt}}.
\end{equation}

Using the scale factors (35)-(37) in the constraint equation (28), the energy density of the fluid is obtained as follows:
\begin{equation}
\rho = 3k^{2}-({c_{4}}^{2}+c_{4}c_{5}+{c_{5}}^{2})e^{-6kt}-\frac{m^{2}-mn+n^{2}}{{c_{3}}^{2}}e^{-2kt-\frac{2}{3}\frac{c_{4}}{k}e^{-3kt}}.
\end{equation}

The deviation-free pressure can be obtained by using (36) and (37) in (29);
\begin{equation}
p=-3k^{2}-({c_{4}}^{2}+c_{4}c_{5}+{c_{5}}^{2})e^{-6kt}.
\end{equation}
Using (37) and (48) in (26), the pressure of the fluid on the $x$, $y$ and $z$ axes are obtained as follows:
\begin{equation}
p_{1}=-3k^{2}-({c_{4}}^{2}+c_{4}c_{5}+{c_{5}}^{2})e^{-6kt}+\frac{n^{2}}{{c_{3}}^{2}}e^{-2kt-\frac{2}{3}\frac{c_{4}}{k}e^{-3kt}},
\end{equation}
\begin{equation}
p_{2}=-3k^{2}-({c_{4}}^{2}+c_{4}c_{5}+{c_{5}}^{2})e^{-6kt}+\frac{m^{2}}{{c_{3}}^{2}}e^{-2kt-\frac{2}{3}\frac{c_{4}}{k}e^{-3kt}},
\end{equation}
\begin{equation}
p_{3}=-3k^{2}-({c_{4}}^{2}+c_{4}c_{5}+{c_{5}}^{2})e^{-6kt}-\frac{mn}{{c_{3}}^{2}}e^{-2kt-\frac{2}{3}\frac{c_{4}}{k}e^{-3kt}}.
\end{equation}

Finally, using the the scale factors in (32), we obtain the heat flow as follows:
\begin{equation}
h_{3}=n(c_{4}-c_{5})e^{-3kt}-m(2c_{4}+c_{5})e^{-3kt}.
\end{equation}

We observe that all the parameters are finite at $t=0$. While the volume of the universe increases continuously, the scale factors may behave non-monotonically for small $t$ values, but eventually all expand monotonically and diverge as $t\rightarrow \infty$. $\Delta$ is null throughout the history of the universe if $c_{4}=c_{5}=0$; otherwise, it decreases monotonically and tends to zero as $t\rightarrow \infty$. Energy density of the fluid may behave non-monotonically for small $t$ values, but eventually tends to $3k^2$ as $t\rightarrow \infty$. The heat flow is null throughout the history of the universe in two cases; i) if $c_{4}=c_{5}(m+n)/(n-2m)$ and ii) if $c_{5}=0$ and $m=n/2$ simultaneously; otherwise, it decreases monotonically and tends to zero as $t\rightarrow \infty$. We also need to examine the late time behavior of $h_{3}/\rho$ and one can observe that it also tends to zero as $t\rightarrow \infty$. Above analysis shows that the model approaches isotropy. The three-Ricci tensors and three-Ricci scalar may behave non-monotonically for small $t$ values, but eventually tend to zero as $t\rightarrow \infty$. Accordingly, the universe eventually evolves to a phase where it is indistinguishably close to a flat and isotropic spatial geometry independent of the three-curvature parameters $m$ and $n$. The behaviors of the directional pressures may differ from each other and may be non-monotonic for small $t$ values, but eventually all of them tend to $-3k^2$ as $t\rightarrow \infty$. Thus the directional EoS parameters ($w_{i}=p_{i}/\rho$), for large $t$ values, are as follows: $w_{1}\cong w_{2}\cong w_{3}\cong -1$. Hence, the fluid we considered isotropizes and exhibits a behavior indistinguishable from that of the conventional vacuum energy ($p_{\rm{vac}}/\rho_{\rm{vac}}=-1$, which is mathematically equivalent to the cosmological constant $\rm{\Lambda}$), in relatively later times of the universe for this model.

Above analysis can be carried out for particular B$\rm{VI_{0}}$, BIII, BV and BI cases by giving appropriate values to the spatial curvature parameters $m$ and $n$. However, as mentioned above, the kinematical properties of the models are independent of the spatial curvature parameters (hence, the type of the metric), but the dynamical quantities of the fluid, the heat flow, and trivially the spatial curvature properties of the models do. Thus, below, we briefly mention only some properties of the models that are particular to the considered Bianchi type.

\paragraph{The case for Bianchi type $\mathbf{\rm{VI_{0}}}$ ($\mathbf{n=m\neq 0}$):}
In this model $h_{3}$ is null if $c_{4}=-2c_{5}$. ${^{3}R^{1}_{1}}={^{3}R^{2}_{2}}$ are null, but ${^{3}R^{3}_{3}}$ and $^{3}R$ are identical and may behave non-monotonically for small $t$ values, but tend to zero as $t\rightarrow\infty$. The directional pressures on the $x$, $y$ axes are identical, and are always higher than the one on the $z$ axis; i.e., $p_{1}=p_{2}>p_{3}$. They may behave non-monotonically for small $t$ values, but eventually all of them tend to $-3k^2$ as $t\rightarrow\infty$.

\paragraph{The case for Bianchi type III ($\mathbf{n=0\neq m}$):}
One can check that if $c_{5}=-2c_{4}$, then the heat flow is null and the model would be equivalent to the one given by Akarsu and Kilinc \cite{AkarsuBIII} for exponential volumetric expansion. ${^{3}R^{2}_{2}}$ is null, but ${^{3}R^{1}_{1}}={^{3}R^{3}_{3}}=\;^3R/2$ and they may behave non-monotonically for small $t$ values, but eventually tend to zero as $t\rightarrow\infty$. The directional pressures on the $x$, $z$ axes are identical, and are always lower than the one on the $y$ axis. They may behave non-monotonically for small $t$ values, but eventually all of them tend to $-3k^2$ as $t\rightarrow\infty$.

\paragraph{The case for Bianchi type V ($\mathbf{n=-m\neq 0}$):}
This model is equivalent to the one obtained by Singh et al. \cite{SinghZeRam} in the presence of a perfect fluid with heat flow for constant mean deceleration parameter with $q=-1$. The heat flow is null if $c_{4}=0$. All the three-Ricci tensors are identical and equal to one third of the three-Ricci scalar. Thus, the curvature of the space is isotropic. They may behave non-monotonically for small $t$ values, but eventually tend to zero as $t\rightarrow\infty$. The directional pressures on the $x$, $y$ and $z$ axes are identical, thus the fluid is isotropic. They may behave non-monotonically for small $t$ values, but eventually all of them tend to $-3k^2$ as $t\rightarrow\infty$.

\paragraph{The case for Bianchi type I ($\mathbf{n=m=0}$):}
This model is the simplest case and equivalent to the one given by Kumar and Singh \cite{KumarSingh} in the presence of a perfect fluid for constant mean deceleration parameter with $q=-1$. The heat flow is null. The three-Ricci tensors and the three-Ricci scalar are all null. Energy density of the fluid increases monotonically and tends to $3k^2$ as $t\rightarrow\infty$. The directional pressures on the $x$, $y$ and $z$ axes are all identical and increase monotonically as the universe evolves and tend to $-3k^2$ as $t\rightarrow\infty$. One may observe that the fluid lies in the phantom region, i.e, $p/\rho<-1$, except in limit.

\subsection{The models for power-law volumetric expansion}
Using the evolution equations (29-31) for the power-law volumetric expansion (34), we obtain the scale factors as follows:
\begin{equation}
A=\frac{a_{1}}{a_{2}a_{3}}t^{l}e^{-\frac{a_{4}+a_{5}}{3l-1}t^{1-3l}},
\end{equation}
\begin{equation}
B=a_{2}t^{l}e^{\frac{a_{5}}{3l-1}t^{1-3l}},
\end{equation}
\begin{equation}
C=a_{3}t^{l}e^{\frac{a_{4}}{3l-1}t^{1-3l}},
\end{equation}
where $a_{2}$ and $a_{3}$ are positive constants and $a_{4}$ and $a_{5}$ are real constants.

The mean Hubble parameter is,
\begin{equation}
H=\frac{l}{t}
\end{equation}
and the directional Hubble parameters on the $x$, $y$ and $z$ axes are, respectively,
\begin{equation}
H_{1}=\frac{l}{t}+(a_{4}+a_{5})t^{-3l},
\end{equation}
\begin{equation}
H_{2}=\frac{l}{t}-a_{5}t^{-3l},
\end{equation}
\begin{equation}
H_{3}=\frac{l}{t}-a_{4}t^{-3l}.
\end{equation}

Using the directional and mean Hubble parameters in (6), the anisotropy of the expansion is obtained as follows:
\begin{equation}
\Delta=\frac{2}{3}\frac{{a_{4}}^{2}+a_{4}a_{5}+{a_{5}}^{2}}{l^{2}}t^{2-6l}.
\end{equation}

Using the scale factor on the $z$ axis (55) in (2) and (3), the three-Ricci tensors and the three-Ricci scalar are obtained as follows:
\begin{equation}
{^{3}R^{1}_{1}}=\frac{m^{2}-mn}{{a_{3}}^{2}}t^{-2l}e^{-\frac{2a_{4}}{3l-1}t^{1-3l}},
\end{equation}
\begin{equation}
{^{3}R^{2}_{2}}=\frac{n^{2}-mn}{{a_{3}}^{2}}t^{-2l}e^{-\frac{2a_{4}}{3l-1}t^{1-3l}},
\end{equation}
\begin{equation}
{^{3}R^{3}_{3}}=\frac{m^{2}+n^{2}}{{a_{3}}^{2}}t^{-2l}e^{-\frac{2a_{4}}{3l-1}t^{1-3l}}
\end{equation}
and
\begin{equation}
^{3}R=\frac{2m^{2}-2mn+2n^{2}}{{a_{3}}^{2}}t^{-2l}e^{-\frac{2a_{4}}{3l-1}t^{1-3l}}.
\end{equation}

Using the scale factors (53)-(55) in the constraint equation (28), the energy density of the fluid is obtained as follows:
\begin{equation}
\rho=\frac{3l^{2}}{t^{2}}-({a_{4}}^{2}+a_{4}a_{5}+{a_{5}}^{2})t^{-6l}-\frac{m^{2}-mn+n^{2}}{{a_{3}}^{2}}t^{-2l}e^{-\frac{2a_{4}}{3l-1}t^{1-3l}}.
\end{equation}

The deviation-free pressure can be obtained by using (54) and (55) in (29);
\begin{equation}
p=\frac{-3l^{2}+2l}{t^{2}}-({a_{4}}^{2}+a_{4}c_{5}+{a_{5}}^{2})t^{-6l}.
\end{equation}
Using (55) and (66) in (26), the pressure of the fluid on the $x$, $y$ and $z$ axes are obtained as follows:
\begin{equation}
p_{1}=\frac{-3l^{2}+2l}{t^{2}}-({a_{4}}^{2}+a_{4}c_{5}+{a_{5}}^{2})t^{-6l}+\frac{n^{2}}{{a_{3}}^{2}}t^{-2l}e^{-\frac{2a_{4}}{3l-1}t^{1-3l}},
\end{equation}
\begin{equation}
p_{2}=\frac{-3l^{2}+2l}{t^{2}}-({a_{4}}^{2}+a_{4}c_{5}+{a_{5}}^{2})t^{-6l}+\frac{m^{2}}{{a_{3}}^{2}}t^{-2l}e^{-\frac{2a_{4}}{3l-1}t^{1-3l}},
\end{equation}
\begin{equation}
p_{3}=\frac{-3l^{2}+2l}{t^{2}}-({a_{4}}^{2}+a_{4}c_{5}+{a_{5}}^{2})t^{-6l}-\frac{mn}{{a_{3}}^{2}}t^{-2l}e^{-\frac{2a_{4}}{3l-1}t^{1-3l}}.
\end{equation}

Finally, using the scale factors in (32), we obtain the heat flow as follows:
\begin{equation}
h_{3}=n(a_{4}-a_{5})t^{-3l}-m(2a_{4}+a_{5})t^{-3l}
\end{equation}

The initial time of the universe is $t=0$, it starts with zero volume and expands indefinitely for all values of $l$. However, because the expansion anisotropy and the three-Ricci scalar contribute the energy density of the fluid negatively, we need to examine (65) to figure out which values of $l$ are convenient for which times of the universe by applying the condition $\rho>0$. Accordingly, if $a_{4}=a_{5}=0$; the models for $l<1$ may represent the relatively earlier times of the universe, the models for $l>1$ may represent the relatively later times of the universe and the model for $l=1$ may represent the entire history of the universe provided that $3{a_{3}}^{2}>{m^{2}-mn+n^{2}}$. Similarly, if ${a_{4}}^{2}+a_{4}a_{5}+{a_{5}}^{2}\neq0$; the models for $0<l\leq 1/3$ may represent the relatively earlier times of the universe, $1/3<l<1$ may represent the intermediate times of the universe and $l>1$ may represent the relatively later times of the universe. Finally the model for $l=1$ may represent the relatively later times of the universe if $3{a_{3}}^{2}>{m^{2}-mn+n^{2}}$, otherwise the intermediate times. We are particularly interested in the models that could eventually evolve to a phase where they are indistinguishably close to a flat and isotropic spatial geometry. Hence, the models for $l>1$ and the model for $l=1$ provided that $3{a_{3}}^{2}>{m^{2}-mn+n^{2}}$ are of our interest, since these are the only models that can represent the late universe and can expand indefinitely without violating the positiveness condition on the energy density of the fluid. One can observe that for these models, expansion anisotropy is null throughout the history of the universe if $a_{4}=a_{5}=0$ and otherwise, it is larger for small $t$ values but decreases as $t$ increases and tends to zero as $t\rightarrow \infty$. Besides, $h_{3}/\rho\rightarrow 0$ tends to zero as $t\rightarrow \infty$. This analysis shows that the models for $l>1$ and the model for $l=1$ provided that $3{a_{3}}^{2}>{m^{2}-mn+n^{2}}$ approaches isotropy. All the three-Ricci tensors and the three-Ricci scalar behave parallel to each other and may behave non-monotonically for small $t$ values but all eventually tend to zero as $t\rightarrow \infty$. Accordingly, the universes that exhibit accelerating power-law volumetric expansion ($l>1$) and the universe that exhibits constant-rate volumetric expansion ($l=1$, provided that $3{a_{3}}^{2}>{m^{2}-mn+n^{2}}$ ), eventually evolve to a phase where they are indistinguishably close to a flat and isotropic spatial geometry independent of the three-curvature parameters $m$ and $n$.

For $l>1$, the energy density of the fluid may behave non-monotonically for relatively small $t$ values, but $\rho\cong 3l^{2}/t^{2}$ for large $t$ values. The directional pressures may behave differently and non-monotonically for small $t$ values, but $p_{1}\cong p_{2}\cong p_{3}\cong (-3l^{2}+2l)/t^{2}$ for large $t$ values. Thus the directional EoS parameters are as follows $w_{1}\cong w_{2}\cong w_{3}\cong -1+\frac{2}{3l}$ for large $t$ values. Accordingly, in the late universe, independently of the spatial curvature parameters, the fluid isotropizes and evolves to an isotropic fluid with an EoS parameter that lies between $-1$ and $-\frac{1}{3}$ according to the value of $l>1$.

For $l=1$ under the condition $3{a_{3}}^{2}>{m^{2}-mn+n^{2}}$, the energy density of the fluid may behave non-monotonically for small $t$ values, but $\rho\cong \frac{3{a_{3}}^{2}-m^{2}+mn-n^{2}}{{a_{3}}^{2}t^{2}}$ for large $t$ values. Similarly, the directional pressures may behave differently and non-monotonically for small $t$ values, but $p_{1}\cong\frac{-1+n^{2}/{a_{3}}^{2}}{t^{2}}$, $p_{2}\cong\frac{-1+m^{2}/{a_{3}}^{2}}{t^{2}}$ and $p_{3}\cong\frac{-1-mn/{a_{3}}^{2}}{t^{2}}$ for large $t$ values. Thus the directional EoS parameters, for large $t$ values, are as follows: $w_{1}\cong -\frac{1}{3u}+\frac{n^{2}}{3u{a_{3}}^{2}}$, $w_{2}\cong -\frac{1}{3u}+\frac{m^{2}}{3u{a_{3}}^{2}}$ and $w_{3}\cong -\frac{1}{3u}-\frac{mn}{3u{a_{3}}^{2}}$, where $u=1-(m^{2}-mn+n^{2})/3{a_{3}}^{2}$. Thus, the fluid does not isotropize even in the very late universe for $l=1$. However, if $a_{3}\gg m$ and $a_{3}\gg n$ simultaneously, then $w_{1}\cong w_{2}\cong w_{3}\cong-\frac{1}{3}$. Hence, also in this model, the fluid may appear to be isotropic.

Note that, while the late time dynamics of the fluid do not exhibit dependence to the spatial curvature parameters in case of $l>1$, they do in case of $l=1$. Thus below, while consdering reduced types, we shall only mention $l=1$ cases for the late time behaviors of the EoS parameters.

\paragraph{The case for Bianchi type $\mathbf{\rm{VI_{0}}}$ ($\mathbf{n=m\neq 0}$):}
In this model, heat flow is null if $a_{4}=-2a_{5}$. ${^{3}R^{1}_{1}}$ and ${^{3}R^{2}_{2}}$ are null, but ${^{3}R^{3}_{3}}$ and the three-Ricci scalar are identical and may behave non-monotonically for small $t$ values, but eventually tend to zero as $t\rightarrow\infty$. The directional pressures on the $x$, $y$ axes are identical, and are always higher than the one on the z axis; i.e., $p_{1}=p_{2}>p_{3}$, both for $l>1$ and $l=1$.

The model for $l=1$ evolves to a phase where it is indistinguishably close to a flat and isotropic spatial geometry provided that $3{a_{3}}^{2}>m^{2}$. The energy density of the fluid may behave non-monotonically for small $t$ values, but $\rho\cong \frac{3{a_{3}}^{2}-m^{2}}{{a_{3}}^{2}t^{2}}$ for relatively large $t$ values. Similarly, the directional pressures may behave differently and non-monotonically for small $t$ values, but $p_{1}=p_{2}\cong\frac{-1+m^{2}/{a_{3}}^{2}}{t^{2}}$ and $p_{3}\cong\frac{-1-m^{2}/{a_{3}}^{2}}{t^{2}}$ for large $t$ values. Thus the directional parameters, for large $t$ values, are as follows: $w_{1}=w_{2}\cong -\frac{1}{3u}+\frac{m^{2}}{3u{a_{3}}^{2}}$ and $w_{3}\cong -\frac{1}{3u}-\frac{m^{2}}{3u{a_{3}}^{2}}$, where $u=1-m^{2}/3{a_{3}}^2$. Thus, the fluid does not isotropize even in the very late universe. However, if $a_{3}\gg m$ then the analysis for the general solution is valid once more.

\paragraph{The case for Bianchi type III ($\mathbf{n=0\neq m}$):}
One can check that if $a_{5}=-2a_{4}$, then the heat flow is null and the model would be equivalent to the one given by Akarsu and Kilinc \cite{AkarsuBIII} for power-law volumetric expansion. ${^{3}R^{2}_{2}}$ is null, but ${^{3}R^{1}_{1}}={^{3}R^{3}_{3}}=\;^3R/2$ and they may behave non-monotonically for small $t$ values, but eventually tend to zero as $t\rightarrow\infty$. The directional pressures on the $x$, $z$ axes are identical, and are always lower than the one on the $y$ axis; $p_{1}=p_{3}<p_{2}$ for both $l>1$ and $l=1$.

The model for $l=1$ evolves to a phase where it is indistinguishably close to a flat and isotropic spatial geometry provided that $3{a_{3}}^{2}>m^{2}$. The energy density of the fluid may behave non-monotonically for small $t$ values, but $\rho\cong \frac{3{a_{3}}^{2}-m^{2}}{{a_{3}}^{2}t^{2}}$ for large $t$ values. Similarly, the directional pressures may behave differently and non-monotonically for small $t$ values, but $p_{1}=p_{3}\cong\frac{-1+n^{2}/{a_{3}}^{2}}{t^{2}}$ and $p_{2}\cong\frac{-1+m^{2}/{a_{3}}^{2}}{t^{2}}$ for large $t$ values. Thus the directional parameters, for large $t$ values, are as follows: $w_{1}=w_{3}\cong -\frac{1}{3u}+\frac{n^{2}}{3u{a_{3}}^{2}}$ and $w_{2}\cong -\frac{1}{3u}+\frac{m^{2}}{3u{a_{3}}^{2}}$, where $u=1-m^{2}/3{a_{3}}^{2}$. Thus, the fluid does not isotropize even in the very late universe. However, if $a_{3}\gg m$ then the preceding analysis is valid once more.

\paragraph{The case for Bianchi type V ($\mathbf{n=-m\neq 0}$):}
This model is equivalent to the one obtained by Singh et al. \cite{SinghZeRam} in the presence of a perfect fluid with heat flow for constant mean deceleration parameter with $q>-1$. The heat flow is null if $a_{4}=0$. All the three-Ricci tensors are identical and equal to one third of the three-Ricci scalar. Thus, the curvature of the space is isotropic. They may behave non-monotonically for small $t$ values, but eventually tend to zero as $t\rightarrow\infty$. The directional pressures on the $x$, $y$ and $z$ axes are identical, thus the fluid is isotropic.

For $l=1$ under the condition ${a_{3}}^{2}>m^{2}$, the energy density of the fluid may behave non-monotonically for small $t$ values, but $\rho\cong \frac{3{a_{3}}^{2}-3m^{2}}{{a_{3}}^{2}t^{2}}$ for large $t$ values. The pressure may behave non-monotonically for relatively small $t$ values, but $p_{1}=p_{2}=p_{3}\cong\frac{-1+m^{2}/{a_{3}}^{2}}{t^{2}}$ for large $t$ values. Thus the EoS parameter, for large $t$ values, is as follows: $w_{1}=w_{2}=w_{3}\cong -\frac{1}{3u}+\frac{m^{2}}{3u{a_{3}}^{2}}$, where $u=1-m^{2}/{a_{3}}^2$.

\paragraph{The case for Bianchi type I ($\mathbf{n=-m\neq 0}$):}
This model is the simplest case and equivalent to the one given by Kumar and Singh \cite{KumarSingh} in the presence of a perfect fluid for constant mean deceleration parameter with $q>-1$. The heat flow is null. The three-Ricci tensors and the three-Ricci scalar are all null. The models for $l>1/3$ and the model for $l=1/3$ under the condition $1/3>{a_{4}}^{2}+a_{4}c_{5}+{a_{5}}^{2}$ satisfy all the conditions for approaching isotropy. For $l>1/3$, the energy density of the fluid may behave non-monotonically for small $t$ values, but $\rho\cong 2l^{2}/t^{2}$ for large $t$ values. The pressure may behave non-monotonically for small $t$ values, but $p_{1}=p_{2}=p_{3}\cong (-3l^{2}+2l)/t^{2}$ for large $t$ values. Thus the EoS parameter, for large $t$ values, is as follows: $w_{1}=w_{2}=w_{3}\cong -1+\frac{2}{3l}$. According to this, the fluid evolves to a fluid with an EoS parameter that lies between $-1$ and $-\frac{1}{3}$ according to the value of $l>1/3$. For $l=1/3$ under the condition $1/3>{a_{4}}^{2}+a_{4}c_{5}+{a_{5}}^{2}$, the energy density and the pressure of the fluid decreases monotonically. Thus the EoS parameter, for large $t$ values, is as follows: $w_{1}=w_{2}=w_{3}\cong 1$.

\section{Conclusion}

We consider the Bianchi type VI spacetime, which can be reduced to Bianchi types $\rm{VI_{0}}$, V, III and I. We derive an energy-momentum tensor under the assumption that the fluid is coupled with the spatial curvature in a way so as to cancel out the terms arise due to the spatial curvature in the evolution equations. This assumption results in an anisotropic pressure in the energy-momentum tensor when the spatial curvature is anisotropic and isotropic pressure when the spatial curvature is isotropic. The energy-momentum tensor we derived secures the evolution of the space, strictly speaking the kinematical properties of the universe, to be independent of the spatial curvature properties. We obtain the exact models for the universes that are expanding indefinitely with constant deceleration parameter. In the presence of the fluid we derived the following: all the models that exhibit accelerating volumetric expansion and the model that exhibits constant-rate volumetric expansion for a particular case, eventually evolve to a phase where they are indistinguishably close to a flat and isotropic spatial geometry independent of whether the fluid is isotropic or anisotropic and the universe possesses isotropic spatial curvature or not. According to this, even if the observed universe is almost isotropic, this does not necessarily imply the isotropy of the fluid affecting the expansion of the universe within the context of general relativity. In particular, we cannot rule out the possible anisotropic character of the inflaton field and dark energy, whose physical natures are not yet understood well.

\begin{center}
\textbf{\textit{Acknowledgments}}
\end{center}
This research was carried out at the Department of Astronomy and Space Sciences, Ege University. \"{O}. Akarsu appreciates the support he received from the Scientific and Technological Research Council of Turkey (T\"{U}B{\.I}TAK) while this research was being carried out. \"{O}. Akarsu acknowledges the support he is presently receiving from the Turkish Academy of Sciences (T\"{U}BA).


\begin{thebibliography}{}
%
%
\bibitem{Ellis06}
Ellis, G.F.R.: Gen. Relativ. Gravit. \textbf{38}, 1003-1015 (2006)
\bibitem{Ellis08}
Ellis, G.F.R.: Cosmological Models. In: Bonometto, S. et al. (eds.) Modern Cosmology, pp. 108-158. Institute of Physics Publishing, Bristol and Philadelphia (2002)
\bibitem{Komatsu10}
Komatsu, E., et al. (WMAP Collaboration): Astrophys. J. Suppl. \textbf{192}, 18 (2011)
\bibitem{Komatsu09}
E. Komatsu, et al. (WMAP Collaboration): Astrophys. J. Suppl. Ser. \textbf{180}, 330-376 (2009)
\bibitem{Spergel07}
D.N. Spergel, et al. (WMAP Collaboration): Astrophys. J. Suppl. Ser. \textbf{170}, 377-408 (2007)
\bibitem{Peiris03}
H.V. Peiris, et al. (WMAP Collaboration): Astrophys. J. Suppl. Ser. \textbf{148}, 213-231 (2003)
\bibitem{Bennett03}
Bennett C.L., et al. (WMAP Collaboration): Astrophys. J. Suppl. Ser. \textbf{148}, 1-27 (2003)
\bibitem{Oliveira}
de Oliveira-Costa, A., et al.: Phys. Rev. D \textbf{69}, 063516 (2004)
\bibitem{Schwarz04}
Schwarz, D.J., et al.: Phys. Rev. Lett. \textbf{93}, 221301 (2004)
\bibitem{Cruz}
Cruz, M., et al.: Astrophys. J. \textbf{655}, 11-20 (2007).
\bibitem{Hoftuft09}
Hoftuft, J., et al.: Astrophys. J. \textbf{699}, 985-989 (2009)
\bibitem{BennettHill}
Bennett, C.L., et al. (WMAP Collaboration): Astrophys. J. Suppl. \textbf{192}, 17 (2011)
\bibitem{Koivisto06}
Koivisto T., Mota, D.F.: Phys. Rev. D \textbf{73}, 083502 (2006)
\bibitem{Barrow97}
Barrow, J.D.: Phys. Rev. D \textbf{55}, 7451 (1997)
\bibitem{Golovnev}
Golovnev, A., et al.: J. Cosmol. Astropart. Phys. \textbf{06}, 009 (2008)
\bibitem{Koivisto08a}
Koivisto T., Mota, D.F.: J. Cosmol. Astropart. Phys. \textbf{08}, 021 (2008)
\bibitem{Rodrigues}
Rodrigues, D.C.: Phys. Rev. D \textbf{77}, 023534 (2008)
\bibitem{Koivisto08b}
Koivisto, M., Mota, D.F.: Astrophys. J., \textbf{679}, 1-5 (2008)
\bibitem{Koivisto08c}
Koivisto, M., Mota, D.F.: J. Cosmol. Astropart. Phys. \textbf{06}, 018 (2008)
\bibitem{AkarsuLRS}
Akarsu, O., Kilinc, C.B.: Gen. Relativ. Gravit. \textbf{42}, 119-140 (2010)
\bibitem{AkarsuSitter}
Akarsu, O., Kilinc C.B.: Astrophys. Space. Sci. \textbf{326}, 315-322 (2010)
\bibitem{AkarsuBIII}
Akarsu, O., Kilinc, C.B.: Gen. Relativ. Gravit. \textbf{42}, 763-775 (2010)
\bibitem{SharifZubair10a}
Sharif, M., Zubair, M.: Astrophys. Space Sci. \textbf{330}, 399-405 (2010)
\bibitem{SharifZubair10b}
Sharif, M., Zubair, M.: Int. J. Mod. Phys. D \textbf{19}, 1957 (2010)
\bibitem{Yadav10a}
Yadav, A.K., Yadav, L.: Int. J. Theor. Phys. \textbf{50}, 218-227 (2011)
\bibitem{Barrow95}
Barrow, J.D.: Phys. Rev. D \textbf{51}, 3113 (1995)
\bibitem{KumarSingh}
Kumar, S., Singh, C.P.: Astrophys. Space. Sci. \textbf{312}, 57-62 (2007)
\bibitem{SinghT}
Singh, T., Chaubey, R.: Astrophys. Space. Sci. \textbf{319}, 149-154 (2009)
\bibitem{Behera10}
Behera, D., Tripathy, S.K., Routray, T.R.: Int. J. Theor. Phys. \textbf{49}, 2569-2581 (2010)
\bibitem{Tiwari09}
Tiwari, R.K.: Astrophys. Space. Sci. \textbf{319}, 85-87 (2009)
\bibitem{SinghJP07}
Singh, J.P., Tiwari, R.K., Shukla, P.: Chin. Phys. Lett. \textbf{24}, 3325 (2007)
\bibitem{Roy85}
Roy, S.R., Singh, J.P., Narain, S.: Astrophys. Space. Sci. \textbf{111}, 389-397 (1985) 
\bibitem{Ram88}
Ram, S.: Int. J. Theor. Phys. \textbf{28}, 97-103 (1989)
\bibitem{Coley90}
Coley, A., Dunn, K.: Astrophys. J. \textbf{348}, 26-32 (1990)
\bibitem{Bali}
Bali, R., Banerjee, R., Banerjee, S.K.: Astrophys. Space. Sci. \textbf{317} 2126 (2008)
\bibitem{Tripathy}
Tripathy, S.K., Behera, D.: Astrophys. Space Sci. \textbf{330}, 191-201 (2010)
\bibitem{Nayak89}
Nayak, B.K., Sahoo, B.K.: General Relativity and Gravitation \textbf{21}, 211-225 (1989)
\bibitem{Roy96}
Roy, S.R., Banerjee, S.K.: General Relativity and Gravitation, \textbf{28}, 27-33 (1996)
\bibitem{SinghZeRam}
Singh, C.P., Zeyauddin, M., Ram, S.: Int. J. Theor. Phys. \textbf{47}, 3162-3170 (2008)
\bibitem{Singh09a}
Singh, C.P.: Gravitation and Cosmology \textbf{15}, 381-390 (2009)
\bibitem{Singh09b}
Singh, C.P.: Astrophys. Space. Sci. \textbf{323}, 197-203 (2009)
\bibitem{Singh07}
Singh, C.P.: Pramana-Journal of Physics \textbf{68}, 707-720 (2007)
\bibitem{Singh08}
Singh, C.P., et al.: Astrophys. Space. Sci. \textbf{315}, 181-189 (2008)
\bibitem{AdhavBansod}
Adhav, K.S., et al.: Astrophys. Space Sci. \textbf{331}, 689-695 (2011)
\bibitem{YadavRahaman}
Yadav, A.K., Rahaman, F., Ray, S.: Int. J. Theor. Phys. \textbf{50}, 871-881 (2011)
\bibitem{PradhanJotania}
Pradhan, A., Jotania, K.: Int. J. Theor. Phys. \textbf{49}, 1719-1738 (2010)
\bibitem{Landau}
Landau, L.D., Lifshitz, E.M.: The Classical Theory of Fields. pp. 82-85. Elsevier, Butterworth Heinemann (2005)
\bibitem{McGlinn}
McGlinn, W.D.: Introduction to Relativity. p. 68. The Johns Hopkins University Press, Baltimore (2003)
\bibitem{CollinsHawking}
Collins, C.B., Hawking, S.W.: Astrophys. J. \textbf{180}, 317-334 (1973)
\bibitem{Gron}
Gr{\o}n, O.: Phys. Rev. D \textbf{32} 2522-2527 (1985)

\end{thebibliography}
\end{document}